\documentclass{PoS}

\title{New Abundances From Very Old Stars}

\ShortTitle{Metal-Poor Stars}

\author{\speaker{Terese Hansen}\\
        Landessternwarte, Heidelberg University\\
        E-mail: \email{thansen@lsw.uni-heidelberg.de}}

\author{C. J. Hansen\\
        Dark Cosmology Center, Copenhagen\\
        E-mail: \email{cjhansen@dark-cosmology.dk}}
\author{N. Christlieb\\
       Landessternwarte, Heidelberg University\\
        E-mail: \email{nchristlieb@lsw.uni-heidelberg.de}}
\author{D. Yong\\
       Research School of Astronomy and Astrophysics, The Australian National
       University, Weston, ACT 2611, Australia\\
       E-mail: \email{yong@mso.anu.edu.au}}
\author{T. C. Beers\\
        Department of Physics and JINA-CEE: Joint Institute for Nuclear Astrophysics -- Center 
        for the Evolution of the Elements, University of Notre Dame, Notre Dame, IN
        46556, USA\\
        E-mail: \email{tbeers@nd.edu}}
\author{J. Andersen\\
        Niels Bohr Institute, University of Copenhagen\\
        E-mail: \email{ja@astro.ku.dk}}
            
\abstract{Metal-poor stars provide the fossil record of Galactic chemical
  evolution and the nucleosynthesis processes that took place at the earliest
  times in the history of our Galaxy. From detailed abundance studies of low
  mass, extremely metal-poor stars ($\mathrm{[Fe/H]} < -3$), we can trace and help
  constrain the formation processes which created the first heavy elements in
  our Galaxy. Here we present the results of a $\sim$25-star homogeneously analysed
  sample of metal-poor candidates from the Hamburg/ESO survey. We have derived
  abundances for a large number of elements ranging from Li to Ba, covering
  production processes from hydrostatic burning to neutron-capture. The sample
  includes some of the most metal-poor stars ($\mathrm{[Fe/H]} < -4$) studied
  to date, and  also a number of stars enhanced in carbon. The so called CEMP
  (carbon enhanced metal-poor) stars, these stars make up $\sim$20\% of the
  stars with $\mathrm{[Fe/H]} < -3$, and  80\% of the stars with
  $\mathrm{[Fe/H]} < -4.5$. The progenitors of CEMP stars are still not fully
  constrained; they could be a result of binary mass transfer or high-mass
  explosive events in the early universe.} 

\FullConference{XIII Nuclei in the Cosmos,\\
		7-11 July, 2014\\
		Debrecen, Hungary }

\begin{document}

\section{Introduction}
The elemental abundances of metal-poor stars (stars with $\mathrm{[Fe/H]}\leq-1$) are widely recognized as
being the footprint of the first stars to form in our Galaxy. The first stars
were presumably very massive due to the lack of cooling agents, resulting in
short lifetimes that ended in a supernova (SN) explosions, thereby enriching
the inter-stellar medium (ISM) with elements heavier than helium. From recent studies of large
samples of metal-poor stars \cite{carollo2012,lee2013,norris2013} it has become
evident that a large fraction of the metal-poor stars are carbon enhanced, the
so-called CEMP stars. The CEMP stars can be divided into several sub-classes,
depending on the level of neutron-capture element abundances they exhibit. The
two main sub-classes of the CEMP stars are the CEMP-$no$ and CEMP-$s$ stars. The CEMP-$no$ stars exhibit no
over-abundances of neutron-capture elements, while the CEMP-$s$ stars exhibit
over-abundances of neutron-capture elements produced by the slow neutron-capture process (s-process). 
CEMP-$s$ stars are generally believed to be the product of mass transfer from a asymptotic giant-branch
(AGB) star companion, which has now evolved into a white dwarf, a picture
supported by radial-velocity monitoring \cite{lucatello2005,hansen2013}. For the
CEMP-$no$ stars, however, the picture is less clear and several progenitors have been
suggested for these stars, including pollution by faint SNe that experienced
extensive mixing and fallback during their explosions
\cite{umeda2003,umeda2005,tominaga2007,tominaga2013,ito2009,ito2013,nomoto2013},
winds from massive, rapidly rotating, mega metal-poor stars, also referred to as ``spinstars''
\cite{hirschi2006,meynet2006,hirschi2007,meynet2010,cescutti2013}, or mass
transfer from an AGB companion \cite{suda2004, masseron2010}. 

\section{Sample}
We have analyzed $\sim$25 metal-poor stars observed with the UVES/VLT
spectrograph, with spectra having an average resolwing power of $R\sim45000$ and signal-to-noise of 40
at 4000\AA. We performed a 1D LTE abundance analysis using the MOOG code with alpha-enhanced
($\mathrm{[\alpha/Fe]}=+0.4$) Kurucz model atmospheres \cite{castellikurucz2003}. Effective temperatures
was determined from the infrared flux method, gravities are from Y$^2$ isochrones \cite{demarque2004} assuming
an age of 10 Gyr and an alpha enhancement of $\mathrm{[\alpha/Fe]}=+0.3$, and
metallicity was derived from Fe~I lines. This analysis procedure was also used
in the Most Metal-Poor Stars project of Yong et al. \cite{yong2013}. Combining
the two datasets results in a total of over 200 homogeneously analyzed stars, a sample that includes some of the most
metal-poor stars known today.

Table \ref{tab1} list the number of stars found in the combined sample for a given metallicity, and
the number of these stars which are carbon enhanced. As can be seen, the
fraction of carbon-enhanced stars rises from about half for stars with
$\mathrm{[Fe/H]}<-2$ to two thirds at $\mathrm{[Fe/H]}<-4$. Table \ref{tab2} shows the stars in the combined
sample divided into the different sub-classes of metal-poor stars,
NMP (for ``normal'' metal-poor), CEMP-$no$, CEMP-$s$, and
CEMP-$r/s$, along with the original definition of the CEMP sub-classes from Beers \&
Christlieb (2005) \cite{beerschristlieb2005}. The majority of the known CEMP stars are of the
CEMP-$s$ variety,  but it has been shown that at the lowest metallicities
($\mathrm{[Fe/H]} < -3.0$) the CEMP-$no$ stars are the dominant sub-class
\cite{aoki2010}. We find a roughly equal number of CEMP-$no$ and CEMP-$s$ stars
in the combined sample, as expected since about half of
the stars in the combined sample have $\mathrm{[Fe/H]}<-3$, where the CEMP-$no$ stars dominate. 

\begin{table}
\center
\begin{tabular}{l|r|r|r|r}
  & $\mathrm{[Fe/H]}<-4$ & $\mathrm{[Fe/H]}<-3$ &
$\mathrm{[Fe/H]}<-2$ & $\mathrm{[Fe/H]}<-1$\\
\hline
Total & 12 & 102 & 208 & 213 \\ 
C-enhanced & 8 & 25 & 55 & 59
\end{tabular}
\caption{Total number and number of carbon-enhanced stars in the combined sample at different metallicities.}
\label{tab1}
\end{table}

\begin{table}
\center
\begin{tabular}{c|c|c|c|c}
Star & NMP & CEMP-$no$ & CEMP-$s$ & CEMP-$r/s$ \\
\hline
Def. &$[$Fe/H$]<-1$ & $[$Fe/H$]<-1$   & $[$Fe/H$]<-1$   &$[$Fe/H$]<-1$ \\
     &              & $[$C/Fe$]>+0.7$  & $[$C/Fe$]>+0.7$  & $[$C/Fe$]>+0.7$ \\
     &              & $[$Ba/Fe$]<+0.0$ & $[$Ba/Fe$]>0.0$ & $[$Ba/Fe$]>0.0$ \\
     &              &                 & $[$Ba/Eu$]>+0.5$ & $0.0<[$Ba/Eu$]<+0.5$\\
\hline
\# & 155 & 24 & 29 & 5 
\end{tabular}
\caption{Definition of the main sub-classes of CEMP stars, and the number of stars in
  each sub-class in the combined sample. Note that we employ $\mathrm{[C/Fe]}>
  +0.7$ to define the CEMP stars, rather than the level of
  $\mathrm{[C/Fe]}>+1.0$ as originally suggested by Beers \& Christlieb (2005).}
\label{tab2}
\end{table}

\subsection{Abundances}
\begin{figure}
\begin{center}
    \includegraphics[width=0.5\textwidth]{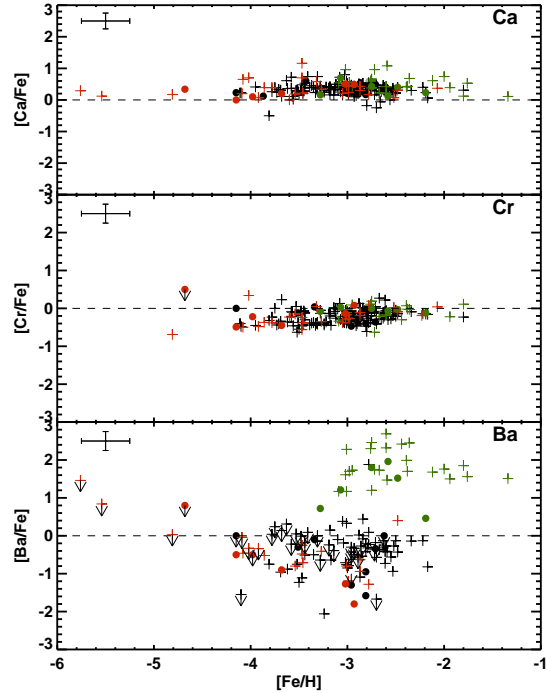}
\caption{Derived abundances for $\mathrm{[Ca/Fe]}$ (top), $\mathrm{[Cr/Fe]}$ (middle), and
  $\mathrm{[Ba/Fe]}$ (bottom), as a function of metallicity. The three
  types of metal-poor stars, NMP (black), CEMP-$no$ (red), and
  CEMP-$s$ (green) in the sample are shown. Dots are from Hansen et al. (2015). and
  the crosses are from Yong et al. (2013). Upper limits are indicated by arrows,
  and the dashed line is the solar value \label{fig1}} 
    \end{center}
\end{figure}

Figure \ref{fig1} shows the derived abundances for the combined sample of Ca,
Cr, and Ba, examples of an alpha-element, iron group-element,
and a neutron-capture element, as a function of metallicity. NMP stars are shown in black, CEMP-$no$ stars in
red, and CEMP-$s$ stars in green. For the alpha-element a small over-abundance is
seen, reflecting the enrichment from one or more core-collapse SNe; also a
very low star-to-star scatter is seen for both the alpha- and iron-group
elements. The abundances of the CEMP-$no$ and CEMP-$s$ stars are
indistinguishable from those of the NMP stars for these two groups of
elements.
 
Examination of the bottom panel of Figure \ref{fig1}, where derived abundances of
the neutron-capture element barium are shown, reveals a quite different
picture. The Ba abundances exhibit a large scatter, indicating more than one
production channel may apply for the neutron-capture elements. The CEMP-$s$
stars are clearly isolated from the rest due to the large over-abundances of
Ba, but there is no clear separation between the Ba abundances of the NMP stars and the CEMP-$no$ stars.

\begin{figure}
\begin{center}
    \includegraphics[width=0.4\textwidth]{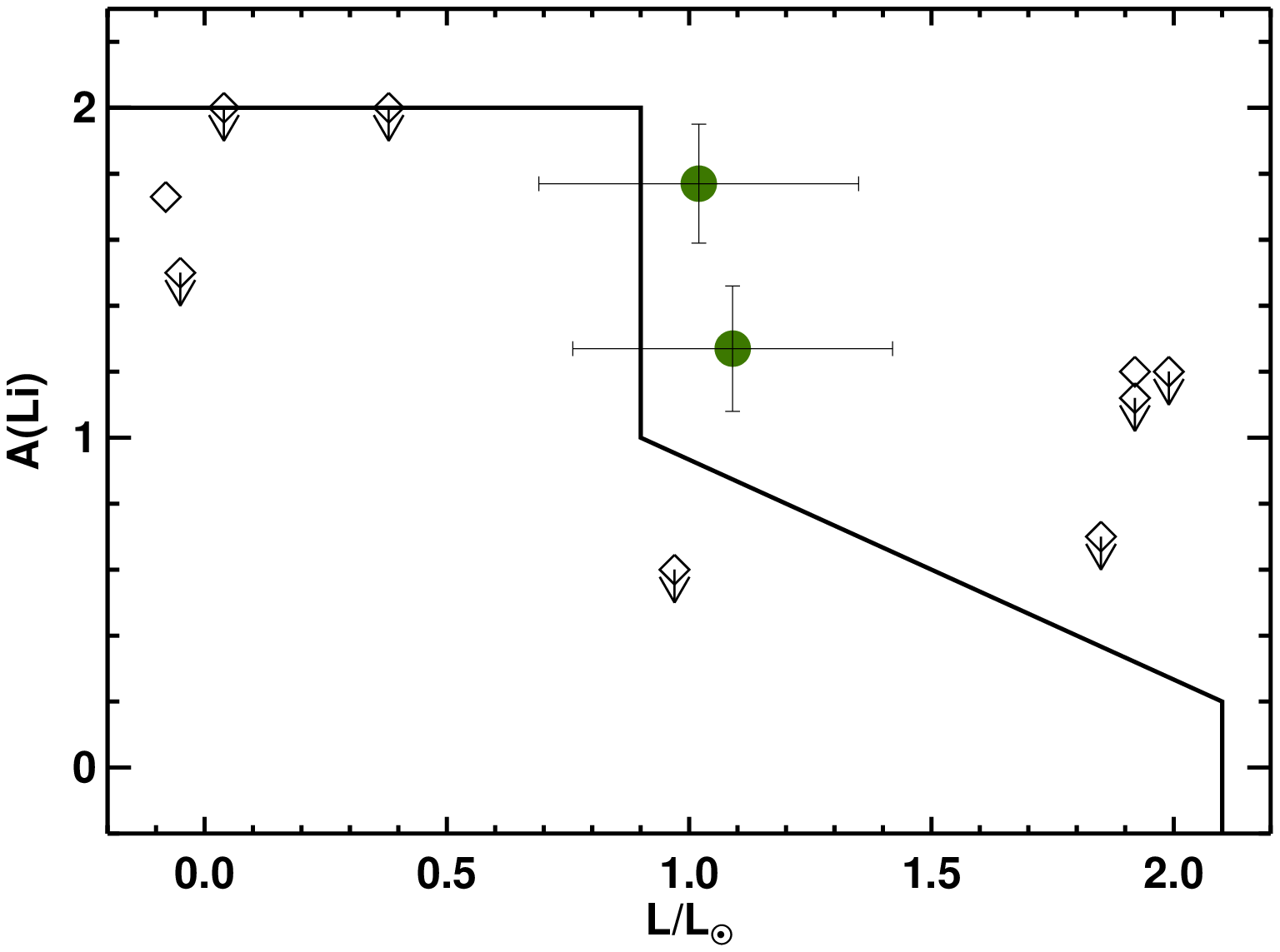}
\caption{Lithium detections for two of the CEMP-$no$ stars in the sample, the
  solid line shows the Spite plateau and the depletion due to stellar
  evolution \cite{hansen2014}. \label{fig2}} 
    \end{center}
\end{figure}

Lithium abundances in metal-poor stars was first explored in the work of Spite and
Spite \cite{spitespite1982}, who discovered the so-called Spite plateau, and concluded that the similar Li abundances for their metal-poor
dwarfs, independent of metallicity, may represent the lithium produced by Big Bang nucleosynthesis. We detect lithium for two of the most metal-poor stars in our sample, 
both of which are ultra metal-poor CEMP-$no$ stars, one with $\mathrm{[Fe/H]}=-4.7$  and one
with $\mathrm{[Fe/H]}=-4.2$ \cite{hansen2014}. Figure \ref{fig2} shows the Li abundances
of these two stars as a function of luminosity, along with Li abundances measured
for other CEMP-$no$ stars taken from Masseron et al. (2012) \cite{masseron2012}. All of the CEMP-$no$ stars have Li
abundances below the Spite plateau, indicating that lithium has been
depleted in these stars. For the giant stars, convection within the star itself
can lead to Li depletion, but for the dwarfs the measured Li abundance should refect that of the ISM
from which the star formed. The Li abundances in CEMP-$no$ stars can be
explained if the progenitors of the CEMP-$no$ stars are massive stars, which
could have destroyed their initial lithium due to internal burning. When this
Li-astrated material mixes with the ISM, the CEMP-$no$ stars that subsequently
form would have overall lower Li abundances \cite{meynet2010,piau2006}.

\subsection{Binary properties}
To asses the binary properties of CEMP-$no$ and CEMP-$s$ stars, we have monitored a
sample of 45 CEMP stars with the Nordic Optical Telescope for a period of $\sim$7
years. The observations have resulted in a dataset of roughly 12 spectra per
star. These have been cross-correlated against a template to
detect any shift in the radial velocity over the seven years of
monitoring. For bright targets, velocity shift down to 300 ms$^{-1}$ can be
detected. Twenty one stars in the sample are CEMP-$no$
stars, of which five show radial-velocity variation. Twelve stars in the
sample are CEMP-$s$ stars, of which ten show variation in their radial velocity over
the seven years of monitoring.

\section{Summary}
We have presented the results of an abundance analysis of $\sim$25 metal-poor
stars, and for the radial-velocity monitoring of another 45 CEMP stars.
At low metallicities a large fraction of the stars are carbon-enhanced. The
two major groups are the CEMP-$s$ and CEMP-$no$ stars; the CEMP-$no$ stars
dominate at metallicities $\mathrm{[Fe/H]}\leq-3.0$. We find abundances for
the alpha elements and iron-peak elements in the CEMP-$s$ and CEMP-$no$ stars that are
indistinguishable from the non carbon enhanced stars in the sample. For the
neutron-capture elements the CEMP-$s$ stars are identified by their high Ba
abundances, where as the CEMP-$no$ and NMP stars have similar abundances.
The lithium abundances found in CEMP-$no$ stars are all below the Spite
plateau, and are suggested to be due to lithium depletion due to burning in
massive first stars and subsequent mixing with primordial gas prior to the
formation of the CEMP-$no$ stars. Results of radial-velocity monitoring demonstrates that, while the pollution from a
binary companion is a good explanation for the abundance abnormalities seen in
the CEMP-$s$ stars, as found in earlier work \cite{lucatello2005}, it is not an acceptable
explanation for the CEMP-$no$ stars. So far  both spinstars and mixing and
fall back SNe are possible progenitors of the CEMP-$no$ stars, but further analysis of larger samples of CEMP-$no$
stars is needed to better constrain these possibilities.

\end{document}